\documentclass{llncs} %

\usepackage{float}
\usepackage{graphicx}

\usepackage[labelfont=bf]{caption}
\usepackage{tabularx}
\usepackage{color}
\usepackage{amsmath} 
\usepackage{mathtools} 
\usepackage{amsfonts} 
\usepackage{amssymb}
\usepackage[normalem]{ulem}

\usepackage{array}
\usepackage{booktabs}
\usepackage{rotating}
\usepackage{multirow}
\usepackage{multicol}
\usepackage{lscape}
\usepackage{subfig}
\usepackage{comment}

\usepackage{wrapfig}
\usepackage{rotating}
\usepackage{epstopdf}

\usepackage{lineno}
\usepackage{soul}
\usepackage{color}
\usepackage{xcolor}
\usepackage{xargs}

\usepackage[export]{adjustbox}

\usepackage{algorithm,algorithmicx,algpseudocode}

\usepackage{color}

\definecolor{brilliantrose}{rgb}{1.0, 0.33, 0.64}

\usepackage[colorlinks=true,linkcolor=blue,citecolor=blue]{hyperref}

\usepackage[colorinlistoftodos,prependcaption,textsize=tiny]{todonotes}
\newcommandx{\ISLEM}[2][1=]{\todo[linecolor=brilliantrose,backgroundcolor=brilliantrose!25,bordercolor=brilliantrose,#1]{#2}}

\usepackage{color}

\definecolor{darkgreen}{rgb}{0.53, 0.66, 0.42}

\usepackage{tikz,xcolor,hyperref}

\definecolor{lime}{HTML}{A6CE39}
\DeclareRobustCommand{\orcidicon}{
	\begin{tikzpicture}
	\draw[lime, fill=lime] (0,0) 
	circle [radius=0.16] 
	node[white] {{\fontfamily{qag}\selectfont \tiny ID}};
	\draw[white, fill=white] (-0.0625,0.095) 
	circle [radius=0.007];
	\end{tikzpicture}
	\hspace{-2mm}
}

\foreach \x in {A, ..., Z}{\expandafter\xdef\csname orcid\x\endcsname{\noexpand\href{https://orcid.org/\csname orcidauthor\x\endcsname}
			{\noexpand\orcidicon}}
}

\begin{document}
\makeatletter
\renewcommand*{\@fnsymbol}[1]{\ensuremath{\ifcase#1\or \dagger\or *\or \ddagger\or
   \mathsection\or \mathparagraph\or \|\or **\or \dagger\dagger
   \or \ddagger\ddagger \else\@ctrerr\fi}}
\makeatother

\title{Recurrent Brain Graph Mapper for Predicting Time-Dependent Brain Graph Evaluation Trajectory}

\titlerunning{Recurrent Brain Graph Mapper}

\author{Alpay Tekin\index{Tekin, Alpay}\inst{1,\ddagger} \and Ahmed Nebli\orcidB{}\index{Nebli, Ahmed}\inst{1,2,\ddagger} \and Islem Rekik\orcidA{}\index{Rekik, Islem}\inst{1}\thanks{ {corresponding author: irekik@itu.edu.tr, \url{http://basira-lab.com}. $\ddagger:$ co-first authors.}} }

\institute{$^{1}$ BASIRA Lab, Faculty of Computer and Informatics, Istanbul Technical University, Istanbul, Turkey \\ $^{2}$ National School of Computer Science (ENSI), University of Manouba, Manouba, Tunisia}

\authorrunning{A. Tekin et al.}

\maketitle       
\begin{abstract}
Several brain disorders can be detected by observing alterations in the brain's structural and functional connectivities. Neurological findings suggest that early diagnosis of brain disorders, such as mild cognitive impairment (MCI), can prevent and even reverse its development into Alzheimer's disease (AD). In this context, recent studies aimed to predict the evolution of brain connectivities over time by proposing machine learning models that work on brain images. However, such an approach is costly and time-consuming. Here, we propose to use brain connectivities as a more efficient alternative for time-dependent brain disorder diagnosis by regarding the brain as instead a large interconnected graph characterizing the interconnectivity scheme between several brain regions. We term our proposed method Recurrent Brain Graph Mapper (RBGM), a novel efficient edge-based \textbf{recurrent graph neural network} that predicts the time-dependent evaluation trajectory of a brain graph from a single baseline. Our RBGM contains a set of \emph{recurrent neural network}-inspired mappers for each time point, where each mapper aims to project the ground-truth brain graph onto its next time point. We leverage the teacher forcing method to boost training and improve the evolved brain graph quality. To maintain the topological consistency between the predicted brain graphs and their corresponding ground-truth brain graphs at each time point, we further integrate a topological loss. We also use $l1$ loss to capture time-dependency and minimize the distance between the brain graph at consecutive time points for regularization. Benchmarks against several variants of RBGM and state-of-the-art methods prove that we can achieve the same accuracy in predicting brain graph evolution more efficiently, paving the way for novel graph neural network architecture and a highly efficient training scheme. Our RBGM code is available at \url{https://github.com/basiralab/RBGM}.

\keywords{Recurrent graph convolution $\cdot$ Transformation layer$\cdot$ Topological loss $\cdot$  Time-dependent graph evolution prediction}

\end{abstract}

\section{Introduction}

Latest neuroscience studies have emphasized the importance of personalized treatments for brain disorders that can significantly improve patient's recovery \cite{lohmeyer2020attitudes}. Brain disorders such as mild cognitive impairment (MCI)  can be easily reversed if diagnosed at an early stage before evolving into irreversible Alzheimer's disease (AD) \cite{stoessl2012neuroimaging}. As such, recent landmark studies \cite{rekik2017joint,gafuroglu2019image,gurler2020foreseeing,nebli2020deep} proposed using the breadth of machine learning to predict brain connectome evolution trajectory. For instance, \cite{rekik2017joint} has proposed a learning-based framework to predict the longitudinal development of cortical surface and white matter fibers. To do so, first, they used multiple atlases to generate a spatially heterogeneous atlas that mimics the cortical surface of the target subject. Second, they predicted spatio-temporal connectivity features from neonatal brains using low-rank tensor completion. Furthermore, \cite{gafuroglu2019image} proposed a deep learning model that jointly classifies and predicts the evolution trajectory of the brain from a single acquisition point. At the baseline, they identified the landmarks, namely regions of interest (ROIs), and utilized supervised and unsupervised learning to predict the evolution trajectory at each landmark.

However, such studies were conducted only on brain images that solely consider local neighborhood connectivities undeniably overseeing the brain global connectivity pattern. To properly diagnose brain connectivity disorders, the brain must be viewed as a large interconnected graph \cite{van2019cross} where each ROI represents a node, and each pairwise connectivity between two ROIs represents an edge. To address this limitation, we set out a more challenging problem, which is predicting the evolution of brain graphs from a baseline observation. In this context, recent works \cite{gurler2020foreseeing,nebli2020deep} leveraged generative adversarial networks (GANs) \cite{goodfellow2014generative}, where they proposed the first graph-based GAN specialized in AD evolution trajectory prediction. 
Namely, GANs \cite{goodfellow2014generative} address the problem of unsupervised learning by training two neural networks: generator and discriminator. The generator takes randomly distributed data as input and generates synthetic data that mimics a real distribution. On the other hand, the discriminator inputs the generated data and predicts whether it is real or fake. Both generator and discriminator compete in an adversarial way. Hence, the overall quality of the generated data improves with training epochs.
\cite{gurler2020foreseeing} used gGAN to learn how to normalize a brain graph with respect to a fixed
connectional brain template (CBT). Their gGAN architecture is made up of a graph normalizer network that learns a high-order representation of each brain in a graph and produces a CBT-based normalized brain graph. Furthermore, \cite{nebli2020deep} proposed EvoGraphNet that connects series of gGANs each specialized in how to construct a brain graph from the predicted brain graph of the previous gGANS in the time-dependent cascade.

Although gGANs were proven to be successful for predicting brain connectome evolution trajectory given a single observation, there is still a considerable amount of computational complexity. Since gGANs contain two graph neural networks (i.e., generator and discriminator), they require a significant amount of computational power and training time. Also, these architectures require a sequential framework where each gGAN has to generate a time point $t_i$ to get the following time point prediction $t_{i+1}$.  

This challenge raises the following question: \emph{can we mimic the prediction power of the above-mentioned sequential gGANs without the need to use highly complex sequential networks?}

According to \cite{mcgrath2011neuroprotection}, brain connectivities do not evolve randomly. Their evolution follows a temporal scheme that aims to satisfy the patient's needs for each given age and health condition. In that regard, recurrent neural networks (RNNs) \cite{connor1994recurrent} are known for their temporal pattern recognition. Therefore, in this paper, we propose to power our model with RNNs. We term our proposed model Recurrent Brain Graph Mapper (RBGM), the \emph{first} framework to predict brain connectome evolution while efficiently reducing complexity and training time consumption. Our model uses a shallow graph convolutional neural network architecture to predict brain disease's evolution trajectory given a baseline graph. To do so, for a given time point, $t_{i}$ each mapper uses the ground-truth from the previous time point $t_{i-1}$ to predict the brain connectivity scheme at the time point $t_{i}$ instead of starting from the initial time point.  To do so, we apply the teacher forcing method \cite{drossos2019language}, which is known for its quick and efficient training recurrent-based models.

We propose preserving the topological consistency between the predicted and ground-truth brain graphs at each time point. To do so, we integrate a topological loss measuring the topological discrepancy between the predicted brain graph and its corresponding ground-truth brain graph. Furthermore, we leverage a $l1$ loss to minimize the sparse distance between two serialized brain graphs to capture time-dependency between two consecutive observations. We also investigate the effect of Kullback-Leibler divergence (i.e., KL-divergence), where we enforce the preservation of node distribution between predicted and ground-truth brain graphs over time. We articulate the main contributions of our work as follows:

\begin{enumerate}

\item  \emph{On a methodological level.} Our proposed RGBM is the first RNN-based geometrical deep learning framework that predicts the time-dependent brain graph evolution trajectory from a single observation.

\item \emph{On a conceptual level.} Our model reduces the complexity of the geometrical deep learning framework by speeding up the training while maintaining similar performances to the state-of-the-art methods. 

\item \emph{On clinical level.} Our RGBM can be used to prevent and reverse the onset of neurological diseases.  

\end{enumerate}

\begin{figure}[h!]
\includegraphics[width=11cm]{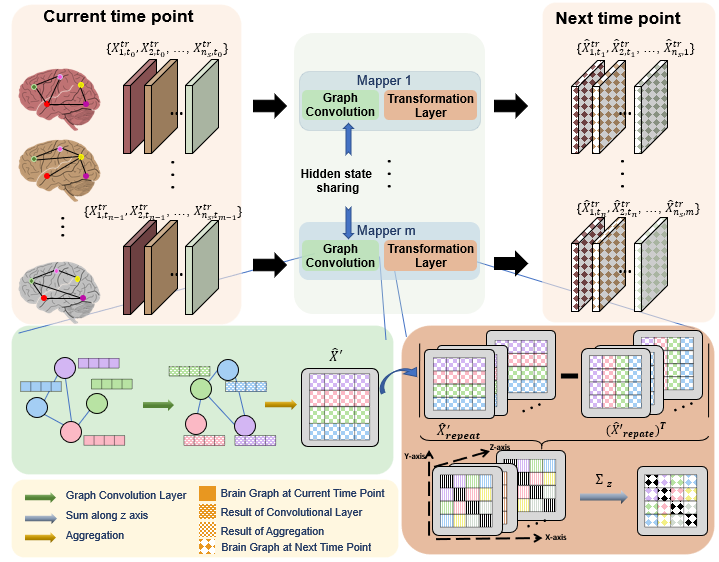} 
\caption{ \emph{Proposed Recurrent Brain Graph Mapper architecture (RBGM) for predicting the evaluation trajectory of brain disease given a single time point.} We develop a mapper that learns how to morph an input at time point $t_{i-1}$ to its next time point $t_{i}$. Given $m$ mappers for $m$ time points, each mapper contains a graph neural network and transformation layer.  The graph neural network takes the input brain graph $\mathbf{X}_{t_{i-1}}^{tr} \in \mathbb{R}^{n_r \times n_r}$ for a given time point $t_{i}$, where $n_r$ is the number of ROIs to learn node embedding $\mathbf{V}^{l} = [\mathbf{v}_{1}^{l}, \mathbf{v}_{2}^{l}, \dots, \mathbf{v}_{n_r}^{l}]^{T}$ that captures the node-to-node relation and visualizes it in vector form. Then a transformation layer takes these node embeddings $\mathbf{V}^{l}$ and computes the pairwise absolute difference to predict the brain graph at the time point $t_{i}$ given by $\mathbf{\hat{X}}_{t_{i}}^{tr} \in \mathbb{R}^{n_r \times n_r}$. First $\mathbf{V}^{l}$ is repeated horizontally $n_{r}$ times to obtain $\mathcal{R} \in \mathbb{R}^{n_r \times n_{r} \times n_{r} }$. Next, we compute the absolute difference between $\mathcal{R}$ and its transpose. Finally, the resulting tensor is sum along $z$-axis to obtain the predicted brain graph $\mathbf{\hat{X}}_{t_{i}}^{tr}$ for the time point $t_{i}$. }
\label{fig:1}
\end{figure}

\section{Proposed Method}

This section introduces the key steps of our RBGM for predicting brain graph evolution from a single observation. \textbf{Table}~\ref{tab:1} displays the mathematical notations that we use throughout our paper. We denote the matrices as boldface capital letters (e.g., $\mathbf{X}$) and scalars as lowercase letters (e.g., $m$). The transpose operator is denoted as $\mathbf{X}^{T}$.

\begin{center}
\begin{table}
\begin{scriptsize}
\caption{\label{tab:1} Mathematical definitions. }
\resizebox{\textwidth}{!}{\begin{tabular}{c c}
 \hline
 \textbf{Mathematical notation}   & \hspace{0.01 cm} \textbf{Definition}  \\
 \hline
 $m$      & number of time points  \\ 
 $n_s$      & number of training subjects  \\
 $m_r$      & number of edges  \\ 
 $n_r$    & number of ROIs in brain \\
 $\mathbf{S}_{i}^{tr}$ & node strength vector of ROI $i$ in the ground-truth brain graph \\
 $\mathbf{\hat{S}}_{i}^{tr}$ & node strength vector of ROI $i$ in the predicted brain graph \\ 
 $\mathbf{X}^{tr}_{t_{i}}$ & training brain graph connectivity matrices $\in \mathbb{R}^{n_{r} \times n_{r}}$ at $t_i$  \\
 $\mathbf{\hat{X}}^{tr}_{t_{i}}$ &  predicted brain graph connectivity matrices $\in\mathbb{R}^{ n_r \times n_r}$ at $t_i$   \\
 $M_i$ & mapper at time point $t_i$ \\
 $\mathcal{L}_{l1}$ & $l_1$ loss \\
 $\mathcal{L}_{TP}$ & Topological loss function \\
 $\lambda_{1}$ & coefficient of $l_1$ loss \\
 $\lambda_{2}$ & coefficient of topological loss \\
 ${V}$ & a set of $n_r$ nodes \\
 ${E}$ & a set of $m_r$ undirected or directed edges\\
 $l$          & index of layer \\
 $\mathcal{N}(i)$       &  the neighborhood containing all the adjacent nodes of node $i$ \\
 $F^l$        &  edge-conditioned filter \\
 $\mathbf{\Theta}^l$ & learnable edge-based parameter for dynamic graph convolution  \\
 ${\mathbf{v}_i}^l$    & node embedding of ROI $i$ at layer $l$ $\in \mathbb{R}^{d_t}$ \\
 $\mathbf{W}^l$        & weight parameter \\
 $\mathbf{b}^l$        & bias term \\
 $\mathbf{\mathcal{R}}$ & horizontally replicated brain connectivity matrix $\in \mathbb{R}^{n_r \times n_r \times n_r} $ \\
 $\mathbf{\mathcal{R}}^{T}$ & transpose of horizontally replicated brain connectivity matrix $\in \mathbb{R}^{n_r \times n_r \times n_r} $\\
 $\mathbf{h}_{i}$ & hidden state matrix at $t_i$ $\in \mathbb{R}^{m_r \times m_r}$\\
 $\mathbf{W}_{ih}$ & input to hidden weight for recurrent filter $\in \mathbb{R}^{1 \times m_r}$ \\
 $\mathbf{W}_{hh}$ & hidden to hidden weight for recurrent filter $\in \mathbb{R}^{m_r \times m_r}$ \\
 $\mathbf{b}_h$    & bias term for recurrent filter $\in \mathbb{R}^{m_r \times m_r}$ \\

 \hline
\end{tabular}}
	\end{scriptsize}
\end{table}
\end{center}

\textbf{Overview of Recurrent Brain Graph Mapper for predicting brain graph evolution trajectory from a single baseline.} Our proposed RBGM is composed of $m$ mappers for $m$ time points, as shown in \textbf{Fig.} \ref{fig:1}. Each mapper can predict a brain graph for a given time point $t_{i}$ using its corresponding ground-truth brain graph at the time point $t_{i-1}$ as an input. The recurrent graph convolution enables each mapper to capture temporal changes in the brain connectivity pattern between consecutive time points. Also, it increases the prediction power of each mapper, hence using fewer convolutional layers. Furthermore, we apply the teacher forcing method \cite{drossos2019language} to quickly and efficiently train our RGBM.

To enhance our method's robustness, we propose using $l1$ loss thanks to its resilience against outliers to enforce the connectivity consistency across time points. Thus, we express the $l1$ loss for each subject $tr$ using the predicted brain graph $\mathbf{\hat{X}}_{t_{i-1}}^{tr}$ from the mapper $M_{i-1}$, and its corresponding ground-truth brain graph at $t_{i}$ as follows:

\begin{gather}
     \mathcal{L}_{l1}(M_{i-1}) = ||\mathbf{\hat{X}}_{t_{i-1}}^{tr} \mathbf{-} \mathbf{X}_{t_{i}}^{tr} ||_{1}
\end{gather}

This acts as a regularizer over time and aligns with the sparse nature of brain connectivity evolution. In addition to the $l1$ loss, we propose a second loss to preserve the topological consistency between predicted brain graphs and their corresponding ground-truth at each time point. To frame the topological loss, we define a node strength vector measuring the topological strength for each node in a given graph. Since our brain graph is fully-connected (each node has a same number of edges), we chose node strength as a centrality measures. We compute the node strength vector by adding the weights of all edges connected to a node of interest. As such,  $\mathbb{\mathbf{S} = [\mathbf{S}_{1}, \mathbf{S}_{2}, ..., \mathbf{S}_{n_{r}}]}^T$ represents the node strengths for all ROIs where $n_r$ is the number of ROIs. The following equation gives the topological loss:
\begin{gather}
    \mathcal{L}_{TP}(\mathbf{S}^{tr}, \mathbf{\hat{S}}^{tr}) =  \frac{1}{n_r} \sum_{i=1}^{n_r} \Big(\mathbf{S}_{i}^{tr} - \mathbf{\hat{S}}_{i}^{tr}\Big)^{2}
\end{gather}
\textbf{The full loss.} We combine the previous losses to train our RGBM as follows:    
\begin{gather}
    \mathcal{L}_{Full} = \sum_{i=1}^{m} \Big(\lambda_{1}\mathcal{L}_{l1}(M_{i-1}) +\lambda_{2} \mathcal{L}_{TP}(\mathbf{S}^{tr},\mathbf{\hat{S}}^{tr}) \Big)
\end{gather}
where $\lambda_{1}$, and $\lambda_{2}$ are hyperparameters adjusting each corresponding loss.

\textbf{The mapper network architecture.} Each mapper $m$ uses our proposed recurrent graph convolution (RGC) function. We leverage the teacher forcing method  \cite{drossos2019language} to speed up training and increase the overall performance in training. Namely, the teacher forcing is a common method that speeds up training and improves the quality of recurrent-based models. It enforces the recurrent model to use ground-truth samples to predict the brain graph at the following time point. According to this method, for a given time point, $t_{i}$ the mapper takes the ground-truth $\mathbf{X}_{t_{i-1}}^{tr}$ from the time point $t_{i-1}$ instead of taking the predicted brain graph $\mathbf{\hat{X}}_{t_{i-1}}^{tr}$ from the preceding mapper $M_{t_{i-1}}$ to make prediction for the time point $t_{i}$ in training phase. Therefore, the need to start from the initial time point is eliminated during training.

\textbf{Dynamic edge-filtered convolution. } Each mapper in our RBGM uses the dynamic graph convolution with edge-conditioned filter proposed by \cite{simonovsky2017dynamic}. Let $G = (V,E)$ is a directed or undirected graph, where $V$ is the set of $n_{r}$ ROIs and $E\in V \times V$ is a set of $m_{r}$ edges between each ROI. Let $l$ be the layer index. For each layer $l \in \{1,2,...,L\}$, $F^{l}: \mathbb{R}^{d_{m}} \mapsto \mathbb{R}^{d_{l}\times d_{l-1}} $ represents a filter-generating network that generate edge weights for the message passing between ROIs $i$ and $j$ given features of $e_{ij}$. $d_m$ and $d_l$ are dimensionality indexes. This operation is expressed as follows:

\begin{gather}
    \mathbf{v}_{i}^{l} = \mathbf{\Theta}^{l}\mathbf{v}_{i}^{l-1} + \frac{1}{|\mathcal{N}(i)|} \bigg(\sum_{j\in \mathcal{N}(i)} F^{l}(\mathbf{e}_{ij};\mathbf{W}^{l})\mathbf{v}_{j}^{l-1} + \mathbf{b}^{l}\bigg),
\end{gather}

where $\mathbf{v}_{i}^{l}$ is the node embedding for the ROI $i$ at layer $l$. $\mathcal{N}(i)$ denotes the neighbors of ROI $i$. $F^{l}$ is the neural network that maps $\mathbb{R}^{d_{m}}$ to $\mathbb{R}^{d_{l}\times d_{l-1}}$ with weights $\mathbf{W}^l$.  $\mathbf{\Theta}^{l}$ is the dynamically generated edge specific weights by $F^l$. The $\mathbf{b}^{l} \in \mathbb{R}^{d_{l}} $ is the bias term. We note that $F^{l}$ can be any type of neural network.

We draw inspiration from the image-based recurrent neural network architecture, which shows outstanding performances on time-series data prediction \cite{cui2019rnn,xu2018automatic}. This type of network can remember the former information and process new events accordingly thanks to its hidden state, which holds the former information (i.e., learned information in the previous layer). Each RNN cells takes two distinct inputs: (i) the input brain graphs from the current time point, and (ii) hidden state value from the brain graphs at the previous time point, then updates the hidden state, which holds the representation of the knowledge from the prior time point.

\begin{figure}[ht!]
\centering
\includegraphics[width=12cm]{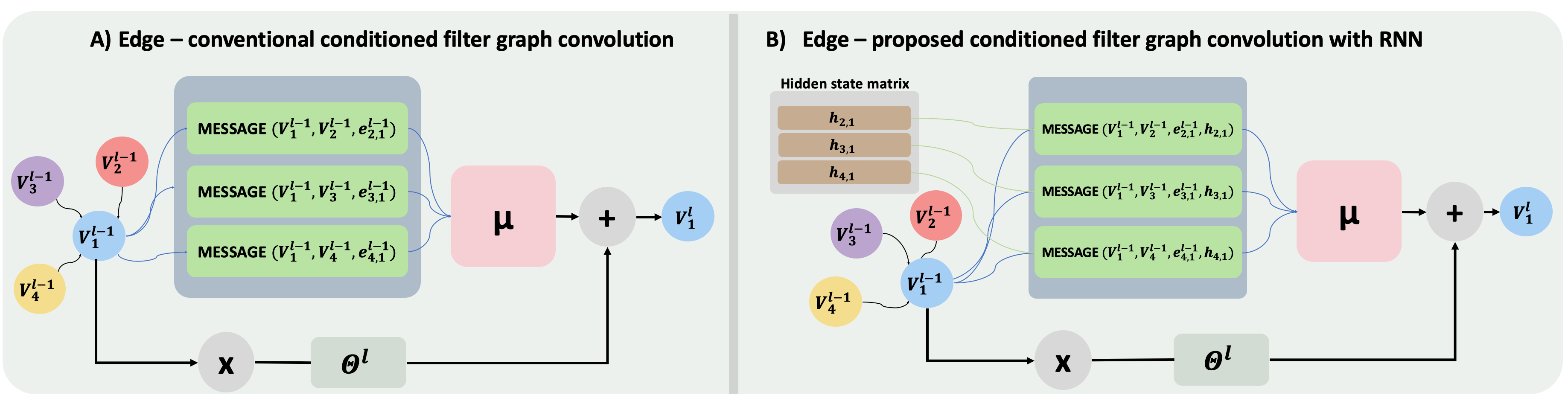}
\caption{ \emph{Illustration of the key differences between the conventional edge-conditioned filter graph generation and our proposed recurrent graph convolution. } \textbf{(A)} \emph{Conventional edge filter for graph convolution.} First, messages are created between ROIs $i$ and its neighbors $\mathcal{N}(i)$. Then, the average of the messages is computed by the mean operation. To inherit the previous layer embedding $\mathbf{V}^{l-1} \in \mathbb{R}^{n_r \times n_r}$, we multiply $\mathbf{V}^{l-1}$ by dynamically generated edge specific weight $\mathbf{\Theta^{l}}$. Finally, $\mathbf{V}^{l}$ is computed by combining the previous layer embedding and the average message passing between ROIs.  \textbf{(B)} \emph{Recurrent graph convolution. } First, the graph recurrent-filter network creates the message between ROIs $i$ and $\mathcal{N}(i)$ by taking hidden state value $h_{ij}$ in contradiction to the conventional edge-conditioned filter graph generation. To inherit previous layer embedding $\mathbf{V}^{l-1} \in \mathbb{R}^{n_r \times n_r}$, we multiply $\mathbf{V}^{l-1}$ by dynamically generated edge specific weight $\mathbf{\Theta^{l}}$. Finally, $\mathbf{V}^{l}$ is computed by combining the previous layer embedding and average message passing between ROIs. }

\label{fig:2}
\end{figure}

\textbf{Proposed graph recurrent-filter.} We propose \emph{the first edge-based recurrent graph neural network} \textbf{Fig.}\ref{fig:2} by re-designing the edge-conditioned filter \cite{simonovsky2017dynamic} in the graph convolution layer as a graph recurrent-filter so that each mapper can capture temporal changes on brain connections over time, as shown in \textbf{Fig.} \ref{fig:2}. Therefore, unlike \cite{nebli2020deep}, the need to start from the initial time point is eliminated during training. Our proposed graph recurrent filter can process past information when generating messages between each ROI to capture temporal changes of brain connectivity. To do so, it takes the set of edges $\mathbf{e} \in \mathbb{R}^{m_r \times 1}$ for a given time point $t_{i}$ and the hidden state matrix from the previous time point $t_{i-1}$ given by $\mathbf{h}^{t_{i-1}} \in \mathbb{R}^{m_r \times m_r}$, which acts as a memory and processes past information. Then it updates the hidden state matrix $\mathbf{h}^{t_{i}} \in \mathbb{R}^{m_r \times m_r}$ for the current time point $t_{i}$. In order to avoid the vanishing gradient problem which makes the network's gradients tend to zero (i.e., hard to learn parameters), we need a function which can bound the gradient and eliminate the risk of divergence during the training. To do so, we use $\tanh$ \cite{shewalkar2019performance} as an activation function in our graph-recurrent filter since it allows the state values to update by bounding in the range of $[-1,1]$ compared to other activation functions such as sigmoid. The equation for our recurrent edge-filtering function $F^{l}(\mathbf{e}^{t_i},\mathbf{h}^{t_{i-1}})$ is expressed as follows:

\begin{gather}
	\mathbf{h}^{t_i} = \tanh( [\mathbf{e}^{t_i},\mathbf{h}^{t_{i-1}}] \odot [\mathbf{W}_{ih},\mathbf{W}_{hh}]^T + \mathbf{b}_{h} ) 	  
\end{gather}

where $\mathbf{W_{ih}} \in \mathbb{R}^{1 \times m_r}$ and $\mathbf{W_{hh}} \in \mathbb{R}^{m_r \times m_r}$ are learnable parameters for input-to-hidden weight and hidden-to-hidden weight respectively. $\mathbf{b_{h}} \in \mathbb{R}^{m_r \times m_r}$ is bias term.

\textbf{The transformation layer architecture.} 
Let  $\mathbf{X}_{t_{i-1}}^{tr} \in \mathbb{R}^{n_{r}\times n_{r}}$ be the input  brain connectivity matrix at a given time point, $t_{i}$ where $n_{r}$ is the number of ROIs. After obtaining the output node embeddings $\mathbf{V}^{l} = [\mathbf{v}_{1}^{l}, \mathbf{v}_{2}^{l},..., \mathbf{v}_{n}^{l}]^{T}$ of RGC layer from a given input $\mathbf{X}_{t_{i-1}}^{tr}$, we construct predicted brain graph $\mathbf{\hat{X}}_{t_{i}}^{tr}$ at $t_{i}$ by computing pairwise absolute difference of learned embeddings \cite{gurbuz2020deep}. To do so, first $\mathbf{V}^{l}$ is replicated with respect to the horizontal axis $n_{r}$ times to obtain $\mathcal{R} \in \mathbb{R}^{n_r\times n_r \times n_r}$. Then, we compute the absolute difference between $\mathbf{\mathcal{R}}$ and its transpose $\mathbf{\mathcal{R}}^{T}$. Finally, the resulting tensor is the sum along $z$-axis to obtain the predicted brain graph $\mathbf{\hat{X}}_{t_{i}}^{tr} \in \mathbb{R}^{n_r \times n_r}$ for the time point $t_{i}$.

\section{Results and Discussion}
\textbf{Evaluation dataset.} We conducted experiments on OASIS-21 longitudinal dataset with $113$ subjects \cite{marcus2010open}. This set contains longitudinal collection of $150$ subjects aged between $60$ to $96$. Each subject's brain scans were acquired $3$ times one year apart. For each subject, we construct a cortical morphological network derived from cortical thickness measure using structural T1-w MRI as proposed in \cite{mahjoub2018brain}. Each cortical hemisphere is parcellated into 35 ROIs using the Desikan-Killiany cortical atlas. We construct our RBGM with PyTorch Geometric library \cite{fey2019fast}.

\textbf{Parameter setting.} In \textbf{Tab}~\ref{tab:2}, we report the mean absolute error between ground-truth and synthesized brain graphs at follow-up time points $t_1$ and $t_2$. In \textbf{Tab}~\ref{tab:3}, we publish the required training time for each comparison method respectively. We set hyperparameters of each mapper as follows: $\lambda_{1}=1$, $\lambda_{2}=10$. We used AdamW \cite{loshchilov2018fixing} optimizer and set the learning rate at $0.0001$ for each mapper. Finally, we trained our model by using $3$-fold cross-validation for $200$ epochs using an NVIDIA Tesla V100 GPU.

\begin{table}

\caption{\label{tab:2} Prediction accuracy of compared methods using MAE at $t_{1}$ and $t_{2}$. }

\resizebox{\textwidth}{!}{\begin{tabular}{c | c | c | c | c}
\hline
 & \multicolumn{2}{c|}{$t_1$} & \multicolumn{2}{c}{$t_2$} \\
 \hline
 \textbf{Method}   
 &  \begin{tabular}{c c}
    \textbf{Mean MAE}   \\
     $\pm$ std
 \end{tabular} 
 & \begin{tabular}{c c}
     \textbf{Best}   \\
     \textbf{MAE}
 \end{tabular}
 &   \begin{tabular}{c c}
     \textbf{Mean MAE}   \\
     $\pm$ std
 \end{tabular} 
 &  \begin{tabular}{c c}
     \textbf{Best}   \\
     \textbf{MAE}
 \end{tabular} \\
 \hline
 
   EvoGraphNet\cite{nebli2020deep}  & $0.05544 \pm 0.01140$ & $0.04555$ & $\mathbf{0.05991 \pm 0.00937}$ & $\mathbf{0.05168}$\\
   
  RBGM (w/KL) &  \hspace{0.2 cm}$0.05585 \pm 0.00349$  \hspace{0.2 cm} &  \hspace{0.2 cm}$0.05341$ \hspace{0.2 cm} &  \hspace{0.2 cm} $0.13509 \pm 0.0098$ \hspace{0.2 cm} & \hspace{0.2 cm} $0.12249$ \hspace{0.2 cm} \\

  RBGM &  $\mathbf{0.04465 \pm  0.00473}$ &  $\mathbf{0.03994}$ &   $0.06228 \pm 0.00315$ &  $0.05870$  \\

 \hline
\end{tabular}}

\end{table}

\begin{table}
\caption{\label{tab:3} Required time for training. }
\setlength{\tabcolsep}{2pt}
\resizebox{\textwidth}{25pt}{\begin{tabular}{c | c | c}
\hline
 \hline
 \textbf{Method}   
 &  \begin{tabular}{c c}
    \textbf{Average Training Time}   \\
 \end{tabular} 
 & \begin{tabular}{c c}
     \textbf{Best Training Time}   \\
 \end{tabular}
 \\
 \hline
EvoGraphNet \cite{nebli2020deep} &  \hspace{0.2 cm}$07:08:33$  \hspace{0.2 cm} &  \hspace{0.2 cm}$02:22:35$ \\
RBGM (w/KL)  & $\mathbf{02:19:40}$ & $\mathbf{00:46:70}$ \\
RBGM & $03:24:08$ & $01:07:50$ \\

 \hline
\end{tabular}}
\end{table}

\textbf{Comparison Method and evaluation.} Due to the lack of RNN-based comparison methods that consider time-dependency for brain graph prediction, we benchmarked our RBGM against some of its variants. We call the first benchmark method: RBGM (w/KL), where we replaced topological loss with KL-divergence loss between the predicted graph at $t_i$ and its corresponding ground-truth brain graph. The KL-divergence minimizes the discrepancy between ground-truth and predicted connectivity weight distribution at each time point $t_i$. The second benchmarking method is against the current state-of-the-art EvoGraphNet \cite{nebli2020deep} in order to assess the power of topological loss and the recurrent graph convolution. In \textbf{Tab}~\ref{tab:2} we report the mean absolute error (MAE) between ground-truth and predicted brain graphs for consecutive time points $t_{1}$ and $t_{2}$ for each comparison method. Our proposed RBGM outperformed baseline methods at $t_{1}$ by achieving both the lowest mean MAE (averaged across the $3$ folds) and the overall best MAE as shown in \textbf{Tab}~\ref{tab:2}. However, for $t_{2}$, EvoGraphNet achieved both the best MAE and mean MAE results. Notably, results show that our RBGM \emph{closely} matches the best results at time point $t_2$ by an error difference of $7 \times 10^{-3}$ in mean MAE yet outperformed EvoGraphNet in time consumption by achieving $46 \%$ less training time.  To the best of our knowledge, such \emph{time/complexity/error} is a delicate compromise to make. Under such a compromisation paradigm, we can fairly judge the \emph{outperformance} of our proposed RBGM is matching state-of-the-art results given less complexity and time consumption.

Overall, our RBGM performs almost similar to the state-of-the-art method EvoGraphNet while speeding up the training and reducing the complexity \textbf{Tab}~\ref{tab:3} since it consists of fewer convolutional layers. However, our RBGM has a few limitations. So far, we have worked on brain graphs where the single edge connects only two ROIs. We aim to generalize our RGBM to handle brain hypergraphs where multiple edges can link two ROIs in our future work. This will enable us to better model and capture the complexity of the brain as a highly interactive network with different topological properties.

\section{Conclusion}

In this paper, we proposed the first edge-based recurrent graph neural network RBGM that uses a novel recurrent graph convolution to predict the brain connectivity evolution trajectory from a single time point. Our architecture contains $m$ number of mappers for $m$ time points. We proposed a time-dependency loss between consecutive time points and a topological loss to preserve topological consistency between predicted and ground-truth brain graphs at the same time point. The results showed that our time-dependent RBGM achieved a similar prediction accuracy compared to the state-of-the-art EvoGraphNet while reducing the training time and complexity. The RBGM is generic and can be used to predict brain graphs for any given time point. In future studies, we aim to generalize our RBGM to using hypergraphs and account for brain hyperconnectivity.

\section{Acknowledgements} 

This work was funded by generous grants from the European H2020 Marie Sklodowska-Curie action (grant no. 101003403, \url{http://basira-lab.com/normnets/}) to I.R. and the Scientific and Technological Research Council of Turkey to I.R. under the TUBITAK 2232 Fellowship for Outstanding Researchers (no. 118C288, \url{http://basira-lab.com/reprime/}). However, all scientific contributions made in this project are owned and approved solely by the authors.

\section{Supplementary material}

We provide three supplementary items for reproducible and open science:

\begin{enumerate}
	\item A 5-mn YouTube video explaining how our framework works on BASIRA YouTube channel at \url{https://youtu.be/QHhvJPyrrSw}.
	\item RBGM code in Python on GitHub at \url{https://github.com/basiralab/RBGM}. 
	\item A GitHub video code demo on BASIRA YouTube channel at \url{https://youtu.be/IkQo9MQHKWo}. 
\end{enumerate}

\bibliography{Biblio3}
\bibliographystyle{splncs}
\end{document}